# Computational Model for Urban Growth Using Socioeconomic Latent Parameters

Piyush Yadav[1, 2], Shamsuddin Ladha[2], Shailesh Deshpande[2] and Edward Curry[1]

[1]Lero-Irish Software Research Centre, National University of Ireland Galway, Ireland
[2]Tata Research Development and Design Centre (TRDDC), TCS Research, Pune, India
`{piyush.yadav,edward.curry}@lero.ie,{shailesh.deshpande,shamshuddin.ladha}@tcs.com`

**Abstract.** Land use land cover changes (LULCC) are generally modeled using multi-scale spatio-temporal variables. Recently, Markov Chain (MC) has been used to model LULCC. However, the model is derived from the proportion of LULCC observed over a given period and it does not account for temporal factors such as macro-economic, socio-economic, etc. In this paper, we present a richer model based on Hidden Markov Model (HMM), grounded in the common knowledge that economic, social and LULCC processes are tightly coupled. We propose a HMM where LULCC classes represent hidden states and temporal factors represent emissions that are conditioned on the hidden states. To our knowledge, HMM has not been used in LULCC models in the past. We further demonstrate its integration with other spatio-temporal models such as Logistic Regression. The integrated model is applied on the LULCC data of Pune district in the state of Maharashtra (India) to predict and visualize urban LULCC over the past 14 years. We observe that the HMM integrated model has improved prediction accuracy as compared to the corresponding MC integrated model.

## 1 Introduction

With an exponential growth in world population, urban growth modelling has become a necessary tool for sustainable and holistic development of any region. A key aspect of urban growth modeling is prediction of land use land cover changes (LULCC) due to various anthropogenic activities. Furthermore, interaction between natural and human factors over large spatio-temporal scale makes LULCC modelling more challenging [1]. Urban growth models usually try to capture influence of growth factors on LULCC to predict the changes (see Fig. 1(a)). These factors are commonly classified as either temporal, or spatial, or spatio-temporal. Temporal factors like National Gross Domestic Product (GDP) change over time but spatially static for a given study area. Spatial factors like Digital Elevation Model (DEM), predominantly, change over space but remain relatively static with respect to time. Spatio-temporal factors such as proximity to the primary roads change over both time and space. Spatial and spatio-temporal factors are also referred to as direct whereas temporal factors are called indirect as, their impact on LULC change cannot be directly quantified. Temporal growth factors can be divided further into supply-side factors (availability of manpower, materials, liquidity, etc.) and demand-side factors (availability of jobs, standard of living, education, health) (see Fig. 1(a)).

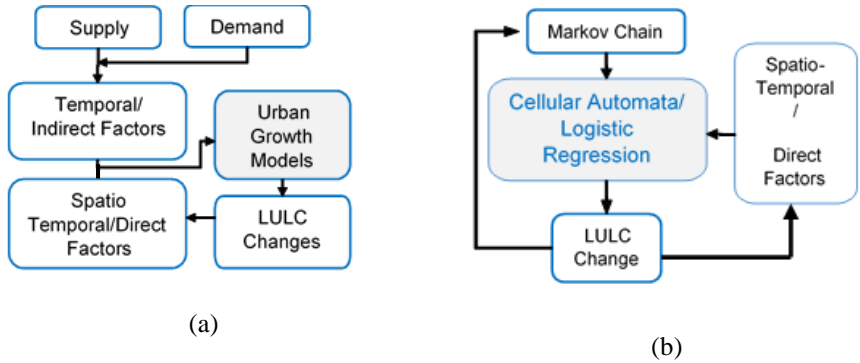

(a) (b)

**Fig. 1. (a)** An urban growth model with multi-scale direct and indirect factors impacting LULC changes **(b)** Schematic of an integrated Markov Chain model

### 1.1 Our Contribution

Our primal contribution is the introduction of Hidden Markov Model (HMM) to incorporate temporal factors in LULC change modeling. We model the underlying temporal factors as Gaussian distributions, conditioned on the hidden states, to learn land cover type transition probabilities. Furthermore, we integrate our model with other spatio-temporal models such as Logistic Regression (LR) to yield richer integrated models than the corresponding MC based integrated models. To our knowledge the introduction of HMM, modeling of temporal factors and integration of HMM based temporal model with other spatio-temporal models has been done for the first time.

The rest of the paper is organized as follows. Section 2 throws light on background and related work. In Section 3, we introduce our integrated model. This entails description of HMM and LR models along with the data used for modeling purposes. We then provide details of the experiments conducted and in-depth analysis of results obtained in Section 4 followed by conclusions and possible future course of directions.

## 2 Background and Related Work

The previous research reports frequent use of Markov Chain (MC) for modeling urban LULC changes. MC is a stochastic model where the states (i.e., land cover or land use classes) correspond to the observable states (events) and these states change over discrete time steps $t = 1, 2, 3, ...$ Given a set of $N$ states, say, $S = \{S_1, S_2, ..., S_N\}$, the MC model is completely specified by the transition probability matrix $\boldsymbol{A} = (\boldsymbol{a_{ij}}) \in \mathbb{R}^{N \times N}$. In order to compute $\boldsymbol{A}$, LULC images of two distinct time instances are considered and the probabilities are computed using the frequency of change from one LULC class to another. MC is a temporal model hence it cannot predict spatial pattern changes.

Recently, MC and other spatio-temporal models have been successfully integrated to model both spatial and temporal patterns of LULC changes (see Fig. 1(b)). The MC model predicts the quantum of growth along with the rate of transfer between different LULC classes (without specifying the exact growth locations), the lattice based spatio-temporal models, e.g. Cellular Automata (CA) and Logistic Regression (LR) [2], effectively model the spatial geographic processes [3]. For example, [4] used MC with CA to integrate open space conservation criteria into the urban growth model. The authors simulated a baseline growth scenario and compared the baseline with a open spaces conservation scenario. However, the MC output for both the scenarios was identical as the open spaces criteria could not be incorporated in MC model. Similarly, [5] [6] [7] [8] have deployed integrated MC-CA models to predict growth scenarios for different regions across the world.

MC is a constrained model that does not account for temporal factors thereby undermining its ability to predict urban growth scenarios. Transition probabilities used in MC model are estimated from the existing land cover changes. This has resulted in computation of future transition probabilities using simple ad hoc approaches in many modeling systems. For example, the future transition probability matrix is derived using a simple power law [9]. More formally, if $a_{11}$ is the transition probability of LC Class 1 over duration $t$, from a reference time $T$, then for time period $2t$ from $T$ new probability value is $a_{11}^2$. Given the complex nature of urban growth, the assumption of persistent rate change is unrealistic. Furthermore, the MC model does not incorporate the effect of other temporal factors such as macro-economics, socio-economic factors, etc. Generally, these factors do not vary spatially for a given urban region. The net effect of these shortcoming of MC is that the temporal factors remain outside the purview of the modeling process. This is true for MC based integrated models such as MC-CA.

## 3 Approach: Model and Data

We propose a new LULCC prediction model that incorporates economic driver factors of urban growth using HMM. We further propose to integrate the outcome of HMM model with other spatio-temporal model such as Logistic Regression (LR) (see Fig. 2). The HMM models diverse temporal (indirect) factors (unlike MC) and predicts the quantum of LULC changes (similar to MC), although more precisely.

It is well known that temporal processes such as LULCC and socio-economic changes are linked and interactive [10]. This warrants a framework that models both the processes satisfactorily. Because of the rich mathematical structure of the Hidden Markov Model (HMM), it encapsulates the two interactive process effectively. We lev-erage this strength and develop a HMM based LULCC model to overcome the limitations of MC model. In the next sections, we describe both the model and the data alongside to facilitate reading.

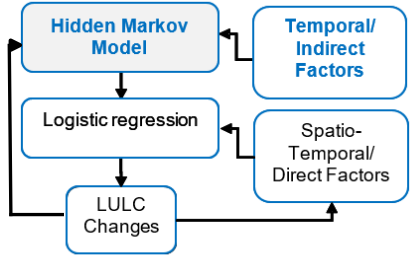 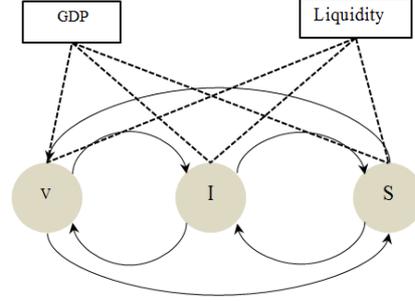

**Fig. 2.** Proposed urban growth model: HMM integrated with Logistic Regression model

**Fig. 3.** A Hidden Markov Model with hidden states (V, I, S) and sample emissions (GDP and Liquidity)

### 3.1 Hidden Markov Model

HMM is a doubly stochastic model [11]. It is defined by a set of $N$ hidden states ($\boldsymbol{S}$) and a set of parameters $\boldsymbol{\theta} = \{\boldsymbol{\pi}, \boldsymbol{A}, \boldsymbol{B}\}$. $\boldsymbol{\pi}$ is a vector of prior probabilities with $\pi_i = P(q_1 = s_i), i = 1 \dots N$, i.e. probability that $s_i$ is the first state ($q_1$) of a state sequence. $\boldsymbol{A}$ is a matrix of state transition probabilities (Eq.(1)). Each $a_{ij} = P(q_{n+1} = s_j | q_n = s_i)$ represents the probability of transitioning from $n^{th}$ state $s_i$ to $(n + 1)^{th}$ state $s_j$. $\boldsymbol{B}$ is a matrix of emission probabilities that characterize the likelihood of seeing an observation $X_n$. Fig. 3 shows a graphical representation of HMM.

In the proposed model, the LULC change is the underlying process that is observed through temporal factors or driver variables. Accordingly, we define the land cover class types as the hidden states. Specifically, the hidden states are defined as Vegetation (V), Impervious surface (I), and Soil (S) land cover class types. $\boldsymbol{S}$ can be used to define hidden states for any land cover or land use class types. The observation $X$ is a vector of temporal factors. These factors are modeled as Gaussian distributions in our model.

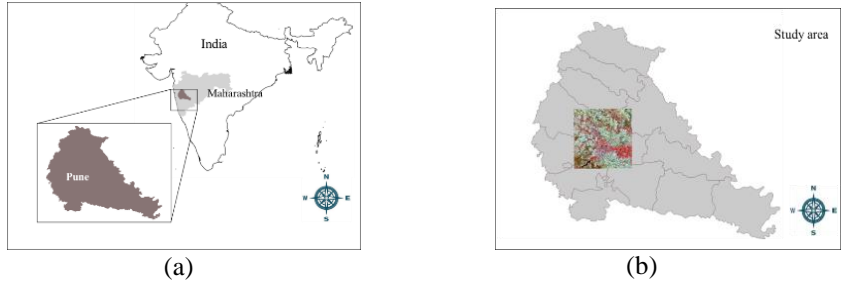

(a)      (b)

**Fig. 4. (a)** Pune district **(b)** Pune district with Landsat image of the area under study

### 3.2 Study Area

Our study area is Pune (see Fig. 4(a-b)), a Tier 1 city situated in the state of Maharashtra, India. It is located 560 m above the sea level and is a part of Deccan plateau region of India. Pune is famous for its Information Technology and Automobile industries and is a hub of various research institutes in India. The study area is core urban Pune that consists of the area under the Pune and Pimpri Chinchwad Municipal Corporation. We have considered 45 sq. km of the city area spanning important features and landmarks which has gone under rapid urbanization in last some decades [12].

**Table 1.** Temporal Growth factors with their type, scale and definition

| Growth Factors | Type[1] | Scale[2] | Definition |
|---|---|---|---|
| Gross Domestic Product | SE | N | Fig. 5 (a) shows the Gross Domestic Produce (GDP) growth rate of India of the past 14 years [13]. Ideally, we should have used Pune's Gross District Domestic Product (GDDP) indicator instead of GDP. However GDDP data was deficient (not available for the entire experimentation period) hence we had to forego the data. |
| Interest Rate Cycle | SE | N | In India, the monetary policy is revised bimonthly. A tight monetary policy affects the overall investment policy which leads to slowdown and vice versa. Fig. 5(b) shows the absolute bimonthly interest rates for the past 14 years [14]. |
| Consumer Price Index Inflation | SE | N | High inflation is one of the major roadblocks in the pathway of economic development. It is evident that low inflation creates developmental investment environment (see Fig. 5(a)) [15]. |
| Gross Fixed Capital Formation | SE | N | GFCF quantifies the amount that the government spends in the capital formation of the country. Capital formations such as infrastructure building, land improvements, machinery and equipment purchases, etc. influence growth. In general, greater the GFCF investment higher is the rate of urbanization (see Fig. 5 (a)) [16]. |
| Urban Population Growth Rate | S | N | It is to be noted that although the growth rate has decreased in the period between the years 2001-2005 and 2006-2010[3] the total urban population has however increased (see Fig. 5(a)). In order to accommodate a higher influx of people, cities are expanding along their outskirts, leading to the growth in urban agglomerate. |
| Per Capita Electricity Consumption | SE | R | Since electricity is one of the major drivers of growth thus per capita consumption of electricity is a good indicator about the demand of electricity in a region. Typically, regions with higher electricity demand grow faster than those with lesser demand. Fig. 5 (c) shows the consumption of electricity in the state of Maharashtra for three major sectors, i.e., Domestic, Industrial, and Agriculture [17]. |
| Road Length Added | SE | R | Roads are one of the basic regional development indicators. Better connectivity of a region helps in better transportation and thus provides impetus to growth by allowing setup of new industrial complexes and other infrastructure services. We used new roads developed in the state of Maharashtra in the past 14 years [17]. |

### 3.3 Temporal Growth Factors for the HMM

There are many temporal factors that impact urban growth, ranging from economic indicators such as Gross Domestic Product, Housing Index, Interest Rate, and Services to social indicators such as Urban Population Growth, Employment Index, and presence of civic amenities.

We have taken multi-scale (i.e., national and regional) socio-economic indicators for the past 14 years (2001-2014). Table 1 lists the growth factors used in our study. This data was collated from different official sources such as National Informatics Centre, The World Bank, Directorate of Economics and Statistics, Planning Department, Government of Maharashtra. Some of the data was obtained from the websites like Trading Economics (see Fig. 5(a-c)). All the growth factors were mean normalized to bring them to a uniform scale between 0 and 1.

---

[1] SE: Socio-economic, S: Social

[2] N: National, R: Regional

[3] Trading Economics Urban Growth Rate

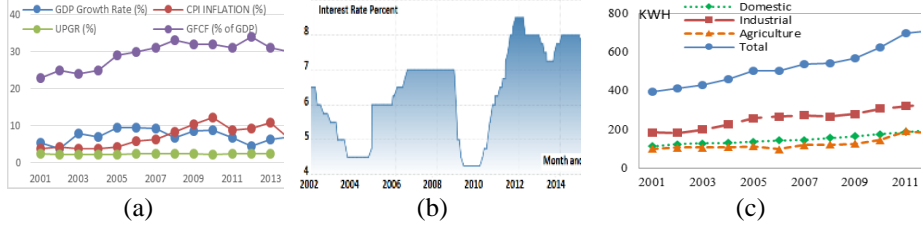

**Fig. 5**. **(a)** GDP growth rate (%), Absolute average CPI Inflation (%), Gross fixed capital formation (% GDP), Urban population growth rate (%) **(b)** Bimonthly interest (repo) rate (%) **(c)** Per capita electricity consumption in kilowatt-hours

### 3.4  Land Use Land Cover (LULC) Data

LULC data is required for HMM hidden states and LR models as an input. Here we describe how LULC data of the study area was systematically derived using remotely sensed data. Images of the study area were acquired from LANDSAT-7 ETM+ sensor from USGS Earth Explorer and Global Visualization Viewer [18]. Table 2 lists the sensor details and Fig. 4(b) shows the corresponding land cover image for the area under study.

**Image Pre-processing**

**Scan Line Correction (SLC):** Since 2003 the Landsat-7 SLC in ETM+ instrument has developed a fault thus creating some black lines in the captured images. To compensate for the lost data, SLC correction was done using sliding window method [19]. A window of $9 \times 9$ pixels was moved over the image and the missing pixels were corrected calculating the mode of the pixels in the window.

**Atmospheric Correction:** Electromagnetic radiation captured by the satellite sensors is affected because of the atmospheric interference such as scattering, dispersion, etc. In each image band of the LANDSAT 7 images, the additive component of atmospheric distortion was corrected by subtracting the digital number (DN) of water pixels in band 4 (infrared band) as it has very low water leaving radiance [20]. In addition, the radiance values of the acquired images were compensated for different solar elevation angles of the each image (see Table 3) [21].

**Table 2.** Landsat-7 Specifics

| Time period | 2001 to 2014 (March to April) |
|---|---|
| **Latitude** | 18.38847838°N - 18.79279909°N |
| **Longitude** | 73.64552005°E - 74.07494971°E |
| **Bands** | 1 to 7 |
| **Resolution** | 30m |
| **Pixels** | $1500 \times 1500$ |

**Table 3.** Atmospheric and Solar Correction

| Atmospheric Correction | Solar Correction |
|---|---|
| $L = L_{min} + \left(\frac{L_{max}}{254} - \frac{L_{min}}{255}\right) \times DN$ where $L$ is the spectral radiance, $L_{min}$ is the minimum spectral radiance, $L_{max}$ is the maximum spectral radiance, and $DN$ is the corrected digital number for each pixel | $\rho_p = \frac{\pi \cdot L_\lambda \cdot d^2}{ESUN_\lambda \cdot \cos\theta_s}$ where $\rho_p$ is unit less planetary reflectance, $L_\lambda$ is spectral radiance at sensor's aperture, $d$ is earth to Sun distance in astronomical units, $ESUN_\lambda$ is mean solar exoatmospheric irradiance, and $\theta_s$ is the solar zenith angle in degrees |

**Image Classification**

Initially, the images were classified into seven broad LULC classes on the basis of the nature of the landscape. These classes were Forest Canopy, Agriculture Area, Residential Area, Industrial Area, Common Open Area, Burnt Grass, Bright Soil, and Water Body. For classification, a labeled set of pixels for each class of interest was collected manually (500 to 3000 samples per class). The feature vector for each pixel consisted of all seven band values. This set was split into train and test datasets. The former was used to train a standard Support Vector Machine (SVM) classifier. The entire image was then classified using the trained classifier. Following this, the seven classes were grouped into three higher level classes namely, Vegetation (V), Impervious Surface (I), and Soil (S).

### 3.5 Logistic Regression and Spatio-Temporal Factors

The class labels of a pixel obtained from image classification are categorical values (e.g., V, I, or S in our case). Logistic regression (LR) is one of the widely used and readily deployable tools available for modeling categorical dependent variables in terms of independent suitability factors. Hence, in our model we have integrated the HMM with an LR model. However, the integration is not restricted to the LR model and the HMM can be readily integrated with other spatio-temporal models. Another purpose of integrating the two models (HMM and LR) is to demonstrate an end to end system for urban growth modelling. Following spatio-temporal driver variables were used as suitability factors for the LR.

**Digital Elevation Model (DEM) and Slope:** The DEM images were acquired from the panchromatic sensors of the satellite CARTOSAT-1 deployed by Indian Space Research Organization (ISRO) [22]. In order to get the DEM of the area under study we merged the two DEM files (coverage: Lat: 18°N-19°N and Long: 73°E-74°E) using QGIS [23] and extracted the DEM for the relevant portion. Slope gradient values were derived from the DEM files using QGIS's Digital Terrain module.

**Proximity to primary roads:** The road data was extracted from the OpenStreetMap of the Pune area. Then the roads of interest were carved out by overlaying road data and Landsat data using QGIS. Using QGIS's Proximity module (based on raster distance), proximity variable was computed.

**Mask:** Water bodies were masked out from the LULC image. The masking was carried out by setting transition probability, in both MC and HMM, from water to any other LC state as 0 and 100% persistence of water bodies over the entire experimentation period.

## 4 EXPERIMENTS AND RESULTS

### 4.1 Hidden Markov Model Experiments

Experiments were conducted on a standard Intel Core i7 machine with Windows 10 and 8 GB RAM. We used Gaussian HMM library of Scikit-learn [24] to conduct HMM experiments and TerrSet [25] for land change modeling experiments using LR. We designed a HMM with the three hidden states (V, I, and S) and temporal factors (see Fig. 5(a-c)) as observations. The first step was to learn the model from the observations. For this purpose, observations from the year 2001 to 2014 were used for model training. HMM was initialized with MC transition probabilities for the year 2001 to 2002 (Table 4- changed to LC class (2002)). MC transition probabilities were obtained using TerrSet's MARKOV module. Initial hidden state ($q_1$) occupancy probability was obtained from the frequency count of LC states of 2001.

**Table 4.** Computed MC transition probabilities for 2001-2002, Learned HMM transition probabilities for 2014, Computed MC transition probabilities for 2014

| Given LC class (2001) | Changed to LC class (2002) | | | Changed to LC class (2014) (HMM) | | | Changed to LC class (2014) (MC) | | |
|---|---|---|---|---|---|---|---|---|---|
| | V | I | S | V | I | S | V | I | S |
| V | 0.7920 | 0.1067 | 0.1013 | 0.8710 | 0.0030 | 0.1260 | 0.6787 | 0.1661 | 0.1542 |
| I | 0.0503 | 0.8996 | 0.0501 | 0.0001 | 0.9610 | 0.0389 | 0.1538 | 0.6484 | 0.1978 |
| S | 0.3058 | 0.1321 | 0.5621 | 0.0020 | 0.1710 | 0.8270 | 0.1372 | 0.0863 | 0.7765 |

HMM was trained using Baum-Welch algorithm [26]. For learning the model the yearly observations were repeated six times (as if they were bi-monthly observations). A stable model was obtained empirically after 50000 iterations with a threshold of less than 0.01. In Fig. 6, Gaussian emission (driver variables) probabilities conditioned on the states (V, I, S) are shown for a sample trained HMM with only three emissions for ease of visualization. Higher values of the observed Emission 1 indicate greater likelihood of the system being in State V. Similarly lower values of Emission1 and Emission2 likely indicate that the system is in the State S.

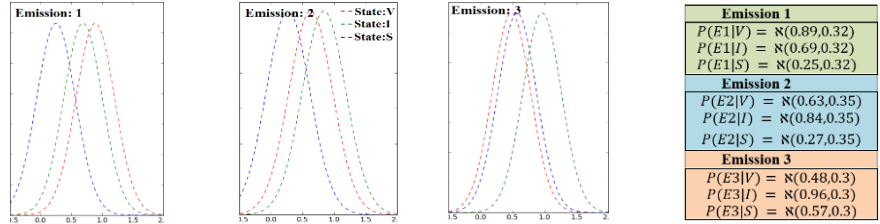

**Fig. 6.** Emission probabilities of a sample HMM with three emissions. P is the probability of seeing an emission given the state and $\aleph(\mu, \sigma)$ is a Gausssian distribution with mean ($\mu$) and standard deviation ($\sigma$)

Comparing the transition probabilities of HMM and MC (Table 4) we find that in MC, persistence of the LC states (i.e. no change in state), including that of the urban state (I), decreases as the prediction period increases. This effect is highlighted in (Table 4 (MC)) with green and orange background colors. While the reduction in persistence of non-urban states (V and S) is typically observed, the persistence of urban states (I) should not reduce significantly over a period of time. This undesired loss of persistence of urban areas is not seen in the HMM transition probabilities (Table 4 (HMM)). Thus HMM transition probabilities are more realistic than those obtained using MC.

### 4.2 Land Change Modeling Experiments

We have used Terrset's Land Change Modeler to conduct land change modeling (LCM) experiments. LC images of the year 2001 and 2009 were used for modeling the spatio-temporal change. Transition sub-models were defined for four LC change types, i.e., V to S, V to I, S to V, and S to I. Water bodies and other protected areas were excluded from the analysis. However, we did not protect existing urban areas from the LCM analysis. Slope gradient and roads layer were used as the primary driver variables for each of the transition sub-models. Slope gradient values ranged between [0-255]. Greater the value of slope for a pixel, lesser is its suitability for urbanization. Hence, the slope gradient variable was transformed with the following negative power function,

$$suitability = \frac{1}{(slope\ gradient)^{0.1}} \quad (1)$$

The transformed variable now represents suitability of a pixel for urbanization; greater the value higher the suitability and vice-versa (Fig. 7(a)). It is evident that the suitability for urbanization is high in areas such as roads, low lying river basin, and around the urbanized areas where the slope gradient is less, and gradually decreases as we move away from the urban areas (Fig. 7(a)). Towards, the south end the suitability drops significantly, as the area has hills and valleys. The discriminatory capabilities of the suitability map were assessed using Cramer's V measure (see Fig. 7(b)). The roads layer data used for LCM is shown in (see Fig. 7(c)). For the purpose of experimentation only primary roads were included.

After processing these spatio-temporal driver variables, each of the four sub-models was built using Logistic Regression (LR). Transition probabilities for the existing urba--nized areas i.e. pixels marked in black are close to zero (see Fig. 8). It is important to note that although the transformed slope gradient suitability of these urbanized areas was high; their transition probability is very small (as expected). In order to predict

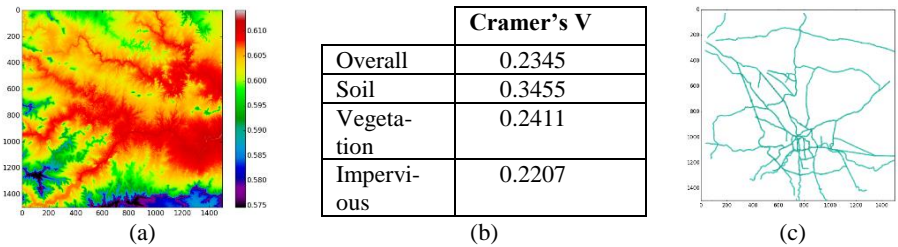

|  | Cramer's V |
|---|---|
| Overall | 0.2345 |
| Soil | 0.3455 |
| Vegetation | 0.2411 |
| Impervious | 0.2207 |

(a)　　　　　　　　　(b)　　　　　　　　　(c)

**Fig. 7. (Left-Right) (a)** Suitability map using transformed slope gradient, **(b)** Cramer's V values for slope gradient suitability, and **(c)** Primary roads map

urban growth, the LR model was integrated with MC (MC-LR) and HMM (HMM-LR) to create two different integrated models.

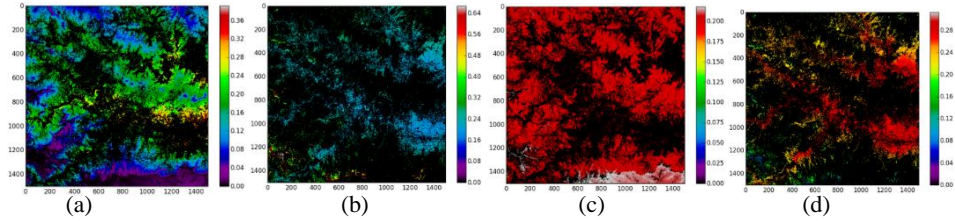

**Fig. 8. (Left-Right)** Heat maps depicting transition probabilities from one state to another **(a)** Soil to Impervious **(b)** Vegetation to Soil **(c)** Soil to Vegetation **(d)** Vegetation to Impervious

The two models were then used to predict changes for the year 2014. The MC-LR model initially estimates the MC transition probabilities for the year 2014 (Table 4 (MC)) and then using these probabilities predicts the changes. On the other hand the HMM-LR model directly uses the learned transition probabilities of the HMM (Table 4 (HMM)) and predicts changes.

Fig. 9 (b) and (c) show predicted LULC changes obtained from HMM-LR and MC-LR integrated models respectively. Visually it is evident that the HMM based predicted image is significantly better, in terms of similarity with the actual classified LC image (Fig. 9(a)), than the MC based predicted image. In many local regions, MC prediction is grossly incorrect with greatly reduced or significantly pronounced urban areas.

Fig. 9(d) shows partial blob analysis on the binary images of urban and non-urban areas. Blobs denote concentrated urban regions. Green blobs are true positives, blue blobs are false negatives, and red blobs are the false positives. HMM-LR false positives are smaller in size and less dense than those of the MC-LR. The HMM output is well balanced and resembles the actual output better. The results obtained using HMM improve prediction accuracy by a large margin. For instance, 11% increment in precision of the persistence of Impervious Surface (I) is observed. Similarly, the precision of Soil (S) class type has jumped up by 26%. However, there is a drop in the precision of Vegetation (V) class type by a marginal 6% (Fig. 9(e)). This is because vegetation cover is an outcome of relatively easy process as compared to S and I. Hence MC-LR captures it with marginal improvement but fail to capture influence of direct variables for S and I which HMM-LR captures efficiently with improved precision and recall.

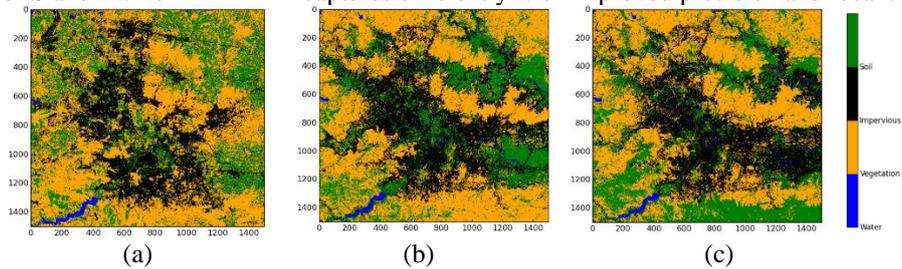

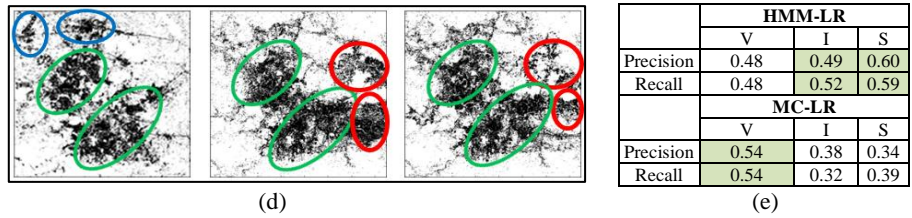

**Fig. 9.** Results for 2014 image **(a)** Actual land cover obtained from classification **(b)** Predicted land cover (HMM-LR) **(c)** Predicted land cover (MC-LR) **(d)** Analysis of urban areas. Left to right: (i) Actual, (ii) MC-LR, (iii) HMM-LR **(e)** Precision and Recall for integrated models

## 5 CONCLUSION

Markov Chain (MC) models are limited in their urban prediction capabilities due to the assumption of constant rate of persistence of land cover class types and inability to model the temporal factors. In this paper, we have proposed a new temporal model using Hidden Markov Model. Our model is richer than MC due to its capability to model urban growth based on important temporal factors. We have demonstrated the correctness of our model over MC by predicting urban growth for Pune city in India.

Our current design of the HMM can model temporal factors limiting its capability to predict of LULC changes. We intend to overcome this limitation by deploying one HMM per pixel location. This would enable modeling of variation of temporal variables over space. In future, it would be interesting to assess the viability of this model with driver variables derived from multiple natural and man-made processes as well.

**Acknowledgment:** The work was supported, in part, by Science Foundation Ireland grant13/RC/2094.